\documentclass[twocolumn,aps,prd,preprintnumbers,nofootinbib]{revtex4}
\usepackage[dvips]{graphicx}
\usepackage{bm,latexsym,amsmath,amssymb,amsfonts}

\usepackage[normalem]{ulem}

\begin{document}

\title{\uline{}Static spacetimes with/without black holes in dynamical Chern-Simons gravity}

\author{Tetsuya Shiromizu${}^{1}$ and Kentaro Tanabe${}^{2}$}
\affiliation{${}^{1}$Department of Physics, Kyoto University, Kyoto 606-8502, Japan}

\affiliation{${}^{2}$Departament de F{\'\i}sica Fonamental, Institut de Ci\`encies del Cosmos, 
Universitat de Barcelona, Mart\'{\i} i Franqu\`es 1, E-08028 Barcelona, Spain}

\begin{abstract}
We show that the static and asymptotically flat black hole spacetime is unique to be 
Schwarzschild spacetime in the dynamical Chern-Simons gravity. In addition, we show that the 
strictly static spacetimes should be the Minkowski spacetime. 
\end{abstract}
\maketitle

\section{Introduction}

Inspired by string theory/loop quantum gravity/particle physics, one may want to consider the dynamical 
Chern-Simons gravity \cite{Alexander:2009tp}. Therein the Chern-Pontryagin term appears as the 
corrections to the Einstein-Hilbert action. Interestingly, this corrections may imply a gravitational 
parity-violation. Therefore, one often discusses some observational consequences from this 
theory \cite{Yagi:2012vf}. 
The Chern-Pontryagin term itself is the topological quantity and then it does not affect the field equation 
in classical level. However if it couples to the scalar fields, it may affect the field equations. 
This is the case of the dynamical Chern-Simons gravity which we will consider here. In this theory, 
the Birkhoff theorem does not hold \cite{Grumiller:2007rv,Yunes:2007ss}. Nevertheless, it turns 
out that the Schwarzschild solution satisfies the field equations \cite{Jackiw:2003pm}. 

In this paper we address if the Schwarzschild solution is unique in the static spacetimes. In the 
Einstein gravity, we know that the staticity of lack hole spacetimes implies the spherical symmetry 
and then the solution 
is unique to be the Schwarzschild spacetime \cite{UT, Israel1968, BM1987, GIS2002}. 
As a consequence, we will see that the key line to prove the uniqueness is not changed in the dynamical 
Chern-Simons gravity. We also see that the strictly static spacetime is the Minkowski 
spacetime as the Einstein theory \cite{Lich, SOS}. By ``strictly static" we mean the presence of 
the hypersurface orthogonal timelike Killing vector in whole spacetime. 

The rest of this paper is organized as follows. In Sec. II, we describe the static spacetimes for 
the dynamical Chern-Simons gravity. In Sec. III, we will prove the static black hole uniqueness and that 
the strictly static spacetime should be the Minkowski spacetime. Finally we give a short summary and 
discussion in Sec. IV. 

\section{Static spacetimes in dynamical Chern-Simons gravity}

The action of the dynamical Chern-Simons gravity is \cite{Alexander:2009tp}
%
\begin{eqnarray}
S & = & \kappa \int d^{4}x{\sqrt {-g}} R +\frac{\alpha}{4}\int d^{4}x{\sqrt {-g}} 
\theta {}^{\ast}R_{\mu\nu\alpha\beta}R^{\mu\nu\alpha\beta} \nonumber \\
& & -\frac{\beta}{2}\int d^{4}x{\sqrt {-g}}(\nabla \theta)^2,
\end{eqnarray}
%
where $\alpha$ and $\beta$ are the coupling constants. $\theta$ is a scalar field. 
${}^{\ast}R_{\mu\nu\alpha\beta}$ is defined by 
%
\begin{eqnarray}
{}^{\ast}R_{\mu\nu\alpha\beta}:=\frac{1}{2}\epsilon_{\alpha\beta}^{~~~\rho\sigma}
R_{\mu\nu\rho\sigma},
\end{eqnarray}
%
where indices $\mu, \nu$ run over $0,1,2,3$, and $\epsilon_{\mu\nu\alpha\beta}$ is the Levi-Civita tensor. 
In this paper we will not consider the potential for $\theta$. Note that our argument in the 
next section does not hold if the potential is. We will briefly discuss the cases with the mass term of 
$\theta$ in Sec. IV. 

The field equations are 
%
\begin{eqnarray}
R_{\mu\nu}-\frac{1}{2}g_{\mu\nu}R+\frac{\alpha}{\kappa}C_{\mu\nu}=\frac{1}{2\kappa}T_{\mu\nu}^\theta
\end{eqnarray}
%
and
%
\begin{eqnarray}
\nabla^2 \theta=\frac{\alpha}{4\beta}R_{\mu\nu\rho\sigma}
{}^\ast R^{\mu\nu\rho\sigma},
\end{eqnarray}
%
where 
%
\begin{eqnarray}
C^{\mu\nu}=\nabla_\sigma \theta \epsilon^{\sigma\delta\alpha (\mu}
\nabla_\alpha R^{\nu)}_\delta
+\nabla_\sigma \nabla_\delta \theta {}^\ast R^{\delta (\mu\nu ) \sigma}
\end{eqnarray}
%
and
%
\begin{eqnarray}
T_{\mu\nu}^\theta=\beta \nabla_\mu \theta \nabla_\nu \theta 
-\frac{\beta}{2}g_{\mu\nu}(\nabla \theta)^2.
\end{eqnarray}
%

In the static spacetimes, the metric is written as 
%
\begin{eqnarray}
ds^2=-V^2(x^i)dt^2+g_{ij}(x^k)dx^i dx^j,\label{metric}
\end{eqnarray}
%
where $g_{ij}$ is the metric of $3$-dimensional 
$t={\rm constant}$ surface $\Sigma$ and indices $i,j$ run over 
$1, 2, 3$. In static spacetimes, the non-trivial components of the Riemann tensor 
are 
%
\begin{eqnarray}
R_{ijkl}={}^{(3)}R_{ijkl} 
\end{eqnarray}
%
and
%
\begin{eqnarray}
R_{0i0j}=VD_i D_j V. 
\end{eqnarray}
%
Then the Ricci tensor are 
%
\begin{eqnarray}
R_{00}=VD^2V 
\end{eqnarray}
%
%
\begin{eqnarray}
R_{ij}={}^{(3)}R_{ij}-\frac{1}{V}D_i D_j V, \label{rij}
\end{eqnarray}
%
and
%
\begin{eqnarray}
R_{0i}=0, 
\end{eqnarray}
%
where ${}^{(3)}R_{ijkl}$, ${}^{(3)}R_{ij}$ and 
$D_i$ are the $3$-dimensional Riemann tensor, Ricci tensor and 
the covariant derivative with respect to $g_{ij}$, respectively. 

In static spacetimes, it is easy to see 
%
\begin{eqnarray}
C_{00}=C_{ij}=T_{0i}^\theta=0, 
\end{eqnarray}
%
%
\begin{eqnarray}
T_{00}^\theta=\frac{\beta}{2}V^2(D \theta)^2,
\end{eqnarray}
%
%
\begin{eqnarray}
T_{ij}^\theta=\beta D_i \theta D_j \theta -\frac{\beta}{2}g_{ij}
(D\theta)^2
\end{eqnarray}
%
and
%
\begin{eqnarray}
C^{0i}& = & \frac{1}{2}D_j \theta \Bigl[ \epsilon^{jkl0}D_lR^i_{~k}+\epsilon^{j0ki}D_k R^0_{~0} 
\nonumber \\
& & +\epsilon^{jk0i}\Bigl( \frac{D_l V}{V}R^l_{~k}-\frac{D_k V}{V}R^0_{~0} \Bigr)\Bigr] \nonumber \\
& & +\frac{1}{2}\epsilon^{ij}_{~~l0} D_j D_k \theta R^{k0l0} \nonumber \\
& & +\frac{1}{4}\epsilon^{0j}_{~~km} D_j D_k \theta R^{kilm}. \label{C0i}
\end{eqnarray}
%
In addition, 
%
\begin{eqnarray}
R_{\mu\nu\alpha\beta}{}^{\ast}R^{\mu\nu\alpha\beta}=0
\end{eqnarray}
%
holds for static spacetimes.

From the field equations we have 
%
\begin{eqnarray}
D^2V=0, \label{d2v0}
\end{eqnarray}
%
%
\begin{eqnarray}
{}^{(3)}R=\frac{2\beta}{\kappa}(D \theta)^2 \label{3ricci}
\end{eqnarray}
%
and 
%
\begin{eqnarray}
D^2\theta +\frac{1}{V}D_i V D^i \theta =0. \label{theta}
\end{eqnarray}
%

\section{Static black hole uniqueness}

In this section, following the argument in Ref. \cite{GIS2002, Rogatko:2002bt}, we will address 
the static black hole uniqueness in the dynamical Chern-Simons gravity. In the static 
spacetimes with the metric of Eq. (\ref{metric}), the event horizon is located at 
$V=0$. At the spatial infinity, we can choose that $V$ goes to unity, that is, $V \to 1$. 

Let us consider the conformal transformation given by 
%
\begin{eqnarray}
\tilde g_{ij}^\pm=\Omega_\pm^2 g_{ij},
\end{eqnarray}
%
where 
%
\begin{eqnarray}
\Omega_\pm=\Bigl( \frac{1\pm V}{2} \Bigr)^2. 
\end{eqnarray}
%
Now we have two manifolds $(\tilde \Sigma_\pm, \tilde g_{ij}^\pm)$. 
In $(\tilde \Sigma_+, \tilde g_{ij}^+)$, it is easy to see that 
the Arnowitt, Deser and Misner(ADM) mass vanishes. On the other hand, from the asymptotic behaviors, 
we can see that the spatial infinity is compactified to a point in 
$(\tilde \Sigma_-, \tilde g_{ij}^-)$. 

At the event horizon, as usual, we impose the regularity of spacetime, that is, the Kretschmann invariant 
is finite there 
%
\begin{eqnarray}
R_{\mu\nu\alpha\beta}R^{\mu\nu \alpha \beta}& = & {}^{(3)}R_{ijkl}{}^{(3)}R^{ijkl}
+\frac{4}{V^2}D_iD_jV D^i D^j V \nonumber \\
& = & {}^{(3)}R_{ijkl}{}^{(3)}R^{ijkl} \nonumber \\
& & +\frac{4}{V^2\rho^2}[k_{ij}k^{ij}+k^2+h^{ij}{\cal D}_i \rho {\cal D}_j \rho],
\end{eqnarray}
%
where $\rho:=(D^iVD_iV)^{-1/2}$, and $h_{ij}, k_{ij}$ and ${\cal D}_i$ are the metric,
 the extrinsic curvature and the covariant derivative of $V={\rm constant}$ surfaces, respectively. 
In the above we used the following relation
%
\begin{eqnarray}
D_i D_j V=\rho^{-1}k_{ij}-\rho^{-2}({\cal D}_i \rho n_j+{\cal D}_j \rho n_i)
-\rho^{-2}kn_in_j,\nonumber \\
\end{eqnarray}
%
where $n^i=\rho D^i V$. Then the regularity on the horizon implies 
%
\begin{eqnarray}
k_{ij}|_{V=0}=0~{\rm and}~{\rho}|_{V=0}=\rho_0={\rm constant.}
\end{eqnarray}
%
Then we see that $\tilde k_{ij}^+|_{V=0}=-\tilde k_{ij}^-|_{V=0}=\rho_0^{-1}\partial_V \Omega_+ h_{ij}|_{V=0}$ 
with $\tilde h_{ij}^+|_{V=0}=\tilde h_{ij}^-|_{V=0}$. Now we can construct the single 
manifold $\tilde \Sigma = \tilde \Sigma^+ \cup \tilde \Sigma^-$ pasting the two 
manifolds along the $V=0$ surface. We can easily see that the Ricci scalar of 
$\tilde g_{ij}^\pm$ is non-negative as 
%
\begin{eqnarray}
\Omega^2_\pm {}^{(3)}\tilde R_\pm={}^{(3)}R \mp 8 \frac{D^2 V}{1 \pm V}=
\frac{2\beta}{\kappa}(D \theta)^2 \geq 0. 
\end{eqnarray}
%
Now we can apply the positive mass theorem \cite{PMT} for $\tilde \Sigma$ and then see that 
it is flat space. This leads us 
%
\begin{eqnarray}
\theta={\rm constant}
\end{eqnarray}
%
and then $C^{0i}=0$. Now the system is reduced to one following the vacuum Einstein equation. 
Thus, the static black hole spacetime should be spherical symmetric 
and then we can see that the spherical symmetric black hole spacetime is the Schwarzschild spacetime 
\cite{Israel1968, BM1987, GIS2002, Rogatko:2002bt}.  

If one knows the detail of the proof in the Einstein theory with non-linear sigma model \cite{Rogatko:2002bt} or 
Einstein-Maxwell-dilaton \cite{EMD}, one may realized that the current proof is basically same with those. 
But, we note that the system here are different from them. 

\section{Strictly static spacetimes}

In this section, we discuss the strictly static spacetime. 

The volume integral of Eq. (\ref{d2v0}) implies that the ADM mass vanishes. Since the scalar 
curvature is non-negative(See Eq. (\ref{3ricci})), the positive mass theorem holds for $t={\rm constant}$ 
surfaces. This implies that $t={\rm constant}$ surfaces are flat space. Then Eq. (\ref{3ricci}) shows us that 
$\theta$ is constant. Thus, Eq. (\ref{d2v0}) becomes 
$\Delta V=0$, where $\Delta$ is the flat Laplacian, and then the regular solution to this is also $V={\rm constant}$. 
Therefore, we could show that the strictly static spacetimes are the Minkowski spacetime in 
the dynamical Chern-Simons gravity. 

\section{Summary and discussion}

In this short paper, we showed that the static black hole should be unique to be the Schwarzschild 
spacetime and the strictly static spacetimes are the Minkowski spacetime. 

We have a comment on the no scalar hair argument by Bekenstein \cite{Bekenstein:1971hc}. 
It is known that we cannot apply the Bekenstein's treatment into the current cases. But, if the scalar 
fields has the mass term, we can do. Then we can show $\theta=0$ and have the same result with the 
current one. 

In this paper we focused on the static spacetimes. One may be interested in the similar issues for 
the stationary spacetimes. This is left for future works. 

\begin{acknowledgments}
We thank the Instituto Superior Tecnico (IST) in Lisbon, where 
this work was initiated during the workshop on ``Strong Gravity beyond GR: from theory to observations". 
We also thank Nicolas Yunes and Paolo Pani for their kind comments. 
TS is supported by Grant-Aid for Scientific Research from Ministry of Education, Science,
Sports and Culture of Japan (No.~21244033). 
KT is supported by a grant for research abroad by JSPS.
\end{acknowledgments}



\end{document}